\documentclass{PoS}

\usepackage{graphicx,amsmath,amssymb}
\usepackage{fontenc, times, mathptmx}
\usepackage{braket}
\usepackage{subcaption}
\usepackage{wrapfig}

\leftline{{\normalsize DESY 20-005} }
\leftline{{\normalsize ADP-20-3/T1113}}
\leftline{{\normalsize LTH-1222}}

\title{Structure functions from the Compton amplitude}

\ShortTitle{Structure functions}

\author{R.~Horsley$^{a}$, Y.~Nakamura$^{b}$,
	 \speaker{H.~Perlt}$^{,c}$, P.~E.~L.~Rakow$^{d}$, G.~Schierholz$^{e}$, 
 	K.~Somfleth$^{f}$, R.~D.~Young$^{f}$, J.~M.~Zanotti$^{f}$\\ \\
 	{\bf \large QCDSF-UKQCD-CSSM collaboration}\\ \\
 	\llap{$^a$} School of Physics and Astronomy,
                    University of Edinburgh,
                    Edinburgh EH9 3FD, UK \\
        \llap{$^b$} RIKEN Advanced Institute for Computational
                    Science, Kobe, Hyogo 650-0047, Japan \\
        \llap{$^c$} Institut f\"ur Theoretische Physik,
                    Universit\"at Leipzig, 04103 Leipzig, Germany \\
        \llap{$^d$} Theoretical Physics Division,
                    Department of Mathematical Sciences,
                    University of Liverpool,
                    Liverpool L69 3BX, UK \\
        \llap{$^e$} Deutsches Elektronen-Synchrotron DESY,
                    22603 Hamburg, Germany \\
        \llap{$^f$} CSSM, Department of Physics, The
        University of Adelaide, Adelaide SA 5005, Australia\\
        E-mail:\\
        \email{holger.perlt@itp.uni-Leipzig.de}}

\abstract{We have initiated a program to compute the Compton amplitude from lattice QCD with the Feynman-Hellman method. 
This amplitude is related to the structure function via a Fredholm integral equation of the first kind. 
It is known that these types
of equations are inherently ill--posed - they are, e.g., extremely sensitive to perturbations of the system.
We discuss two methods which are candidates to handle these problems: the model free
inversion based on singular value decomposition and one Bayesian type approach.
We apply the Bayesian method to currently available lattice data for 
the Compton amplitude.}

\FullConference{37th International Symposium on Lattice Field Theory - Lattice2019\\
		16-22 June 2019\\
		Wuhan, China}

\begin{document}

\section{Introduction}

The determination of the hadronic structure from first principles belongs to the key investigation 
topics in lattice QCD. Central to our understanding of hadron structure are the structure functions which
describe the distribution
of quarks and gluons inside hadrons. In the last years some promising approaches have been proposed, 
among them the calculation of the quasi particle distribution functions (for a review see~\cite{Cichy:2018mum}).
Our group has initiated a program to compute the structure functions from the forward Compton amplitude
of the nucleon~\cite{Chambers:2017dov, Young:2019}. A central motivation for this is to overcome the
issues of renormalization, operator mixing and the restriction to light-cone operators.

The starting point is the forward Compton amplitude of the nucleon~\cite{dis},
\begin{equation}
T_{\mu\nu}(p,q) = \rho_{\lambda \lambda^\prime}\! \int\! {\rm d}^4\!x\, {\rm e}^{iq\cdot x}  \langle p,\lambda^\prime |T J_\mu(x) J_\nu(0)|p,\lambda\rangle \,, 
\label{prod}
\end{equation}
which involves the time ordered product of electromagnetic currents sandwiched between nucleon states of 
momentum $p$ and polarization $\lambda$, where $q$ is the momentum of the virtual photon 
and $\rho$ is the polarization density matrix. In view of our investigation below we consider only the 
unpolarized structure functions. In the unphysical region ($|p\cdot q| < q^2/2 $)
the relation of $T_{\mu\nu}(p,q)$ to the structure functions
$F_1(x,q^2), F_2(x,q^2)$ is given by~\cite{dis}
\begin{eqnarray}
T_{\mu\nu}(p,q) &=& \left(\delta_{\mu\nu}-\frac{q_\mu q_\nu}{q^2}\right)\, 4\omega \int_0^1 dx\, \frac{\omega x}{1-(\omega x)^2}\, F_1(x,q^2)\nonumber \\[0.5em]
&+& \left(p_\mu-\frac{p\cdot q}{q^2}q_\mu\right)\left(p_\nu-\frac{p\cdot q}{q^2}q_\nu\right)\, \frac{8\omega}{2p\cdot q} \int_0^1 dx\, \frac{1}{1-(\omega x)^2}\, F_2(x,q^2)\,,
\label{opes2}
\end{eqnarray}
with $\omega = 2p\cdot q/q^2$, discarding the subtraction term~\cite{Chambers:2017dov}. To simplify the numerical calculation, we may 
choose $\mu = \nu = 3$ and $p_3 = q_3 = q_4 = 0$. We then have
\begin{equation}
T_{33}(p,q) =  4\omega \int_0^1 dx\, \frac{\omega x}{1-(\omega x)^2} F_1(x,q^2) \equiv 
 \int_0^1 dx\, K(x,\omega)\, F_1(x,q^2)\,.
\label{opess2}
\end{equation}
The matrix element $T_{33}(p,q)$ can be computed most efficiently by a simple extension of the 
Feynman--Hellmann method~\cite{Chambers:2017dov,Horsley:2012pz}.

Performing a Taylor expansion of (\ref{opess2}) leads to  a simple relation between 
the moments $t_j=\int_0^1\,dx\,x^j\,F_1(x)$ of the structure function and the $\omega$--dependent
Compton amplitude
\begin{equation}
T_{33}(\omega) = 4\, \big(\omega^2 \,t_1 + \omega^4 \,t_3 + \cdots + \omega^{2M} \,t_{2M-1} + \dots \big) \,.
\label{poly}
\end{equation}
From these we then determine the moments of the parton distributions 
$\mu_j$ from $t_j \sim \mu_j/2$ neglecting logs and terms $O(1/q^2)$.

\section{Problems and solutions of the Fredholm integral equation}

Formula (\ref{opess2}) is the basic relation for our investigation. It tells us how to extract the structure
function $F_1(x,q^2)$ given that we have available lattice data for the Compton amplitude $T_{33}(p,q)$. Unfortunately,
it is a Fredholm integral equation of the first kind. Those equations are known to be ill--posed.  E.g., they are
extremely sensitive to very small perturbations of the data~\cite{Hansen} -- 
in our case to the lattice results of $T_{33}(p,q)$. Additionally, the solutions are not guaranteed to be unique.
There is no general solution method available.
If one finds a successful numerical  strategy at all it depends always on the specific kernel $K$.
Therefore, a careful study of possible  approaches 
is needed. An analogous problem arises in the reconstruction of Ioffe time pseudo particle distribution 
functions (pdf) and was 
investigated in great detail in~\cite{Karpie:2019eiq}.

In order to test some possible numerical methods we generate mock data for the Compton amplitude.
%from phenomenological particle distribution functions based on experiments~\cite{Martin:2009iq}. 
As an example
we choose a valence type up quark distribution 
\begin{equation}
x\,p^{\rm ref}_{u_v}(x) = 5.107 \, x^{0.8}\,(1-x)^3
      \label{pref}
\end{equation}
chosen to satisfy the momentum sum rule
\begin{equation}
 \int_0^1 \,{\rm dx} \,x\,p^{\rm ref}_{u_v}(x) = 1/3\,.
\end{equation}
 
This function is then used to generate the $T_{33}$ data via (\ref{opess2}) and to compare 
with the results of our tested inversion algorithms.

\begin{figure}[b]
  \begin{subfigure}[t]{0.47\textwidth}
    \includegraphics[width=\textwidth]{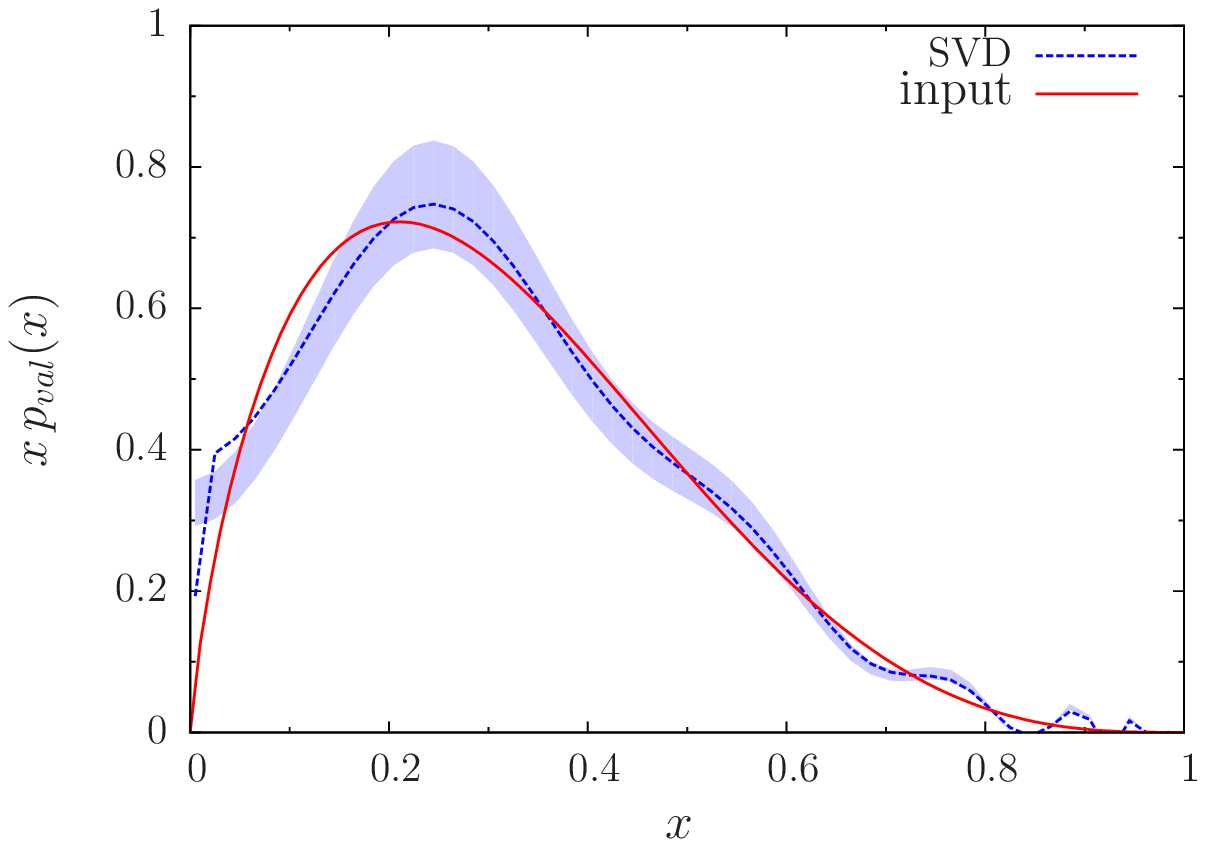}
  \end{subfigure}
  \hspace{0.06\textwidth}
  \begin{subfigure}[t]{0.47\textwidth}
    \includegraphics[width=\textwidth]{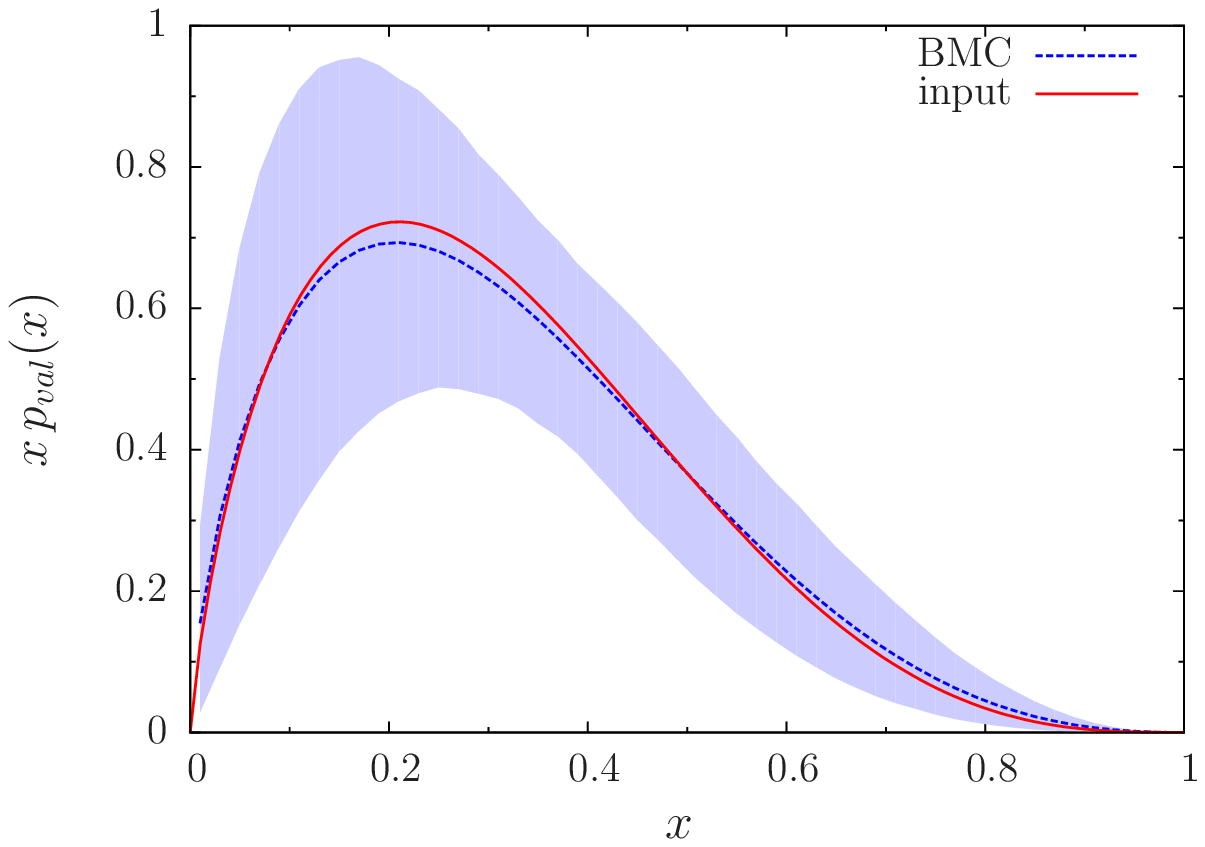}
  \label{fig:T33dat1}
  \end{subfigure}
\vspace{-5mm}
\caption{Left: The pdf as
  obtained from the SVD. The red curve is $x\,p^{\rm ref}_{u_v}(x)$ (\protect\ref{pref}). The blue shadowed area shows the 
  variation of the result $x\,p^{\rm SVD}_{u_v}(x)$ due to a $\pm 10 \%$  variation of $T_{33}$. Right: The pdf from
  the BMC approach. The red curve is again the input (\protect\ref{pref}). The shadowed area is the 
  $68 \%$ quantile.}
  \label{fig:T33dat}
\end{figure}
The numerical inversion requires a discretization of (\ref{opess2})
\begin{equation}
 T_{33}(\omega_i) = \sum_{j=1}^N \, K(x_j,\omega_i)\, p(x_j)
 \leftrightarrow T_{33,i}= \sum_{j=1}^N \, K_{ji}\, p_j \,, i = 1 \dots M \,,
 \label{FId}
\end{equation}
where in general we have $N \neq M$.

One basic method to solve (\ref{FId}) for the $p_j$ is the singular value decomposition (SVD)~\cite{nr}.
It has the advantage that one does not need to make any further input assumptions about the expected form of the 
wanted $p(x)$. On the other hand there is a certain freedom in omitting small singular values. Additionally, 
using our kernel $K(x,\omega)$ (\ref{opess2}) we have cancellations of very large numbers which increases with the number of included
singular values. This demands very precise lattice data in order to get meaningful results. The result of
the inversion $x\,p^{\rm SVD}_{u_v}(x)$ is shown in the left panel of Fig. \ref{fig:T33dat} for $N=50$,
$M=10$ and $0< \omega < 1$. One recognizes 
a trend around the the input distribution with some small oscillations . The integral using the mean value is
%%\begin{equation}
$ \int_0^1\,{\rm dx} x\,p^{\rm SVD}_{u_v}(x) \approx 0.33\,.$
%%\end{equation}

An alternative approach uses some prior model information concerning the distribution and 
tries to refine it according to the available data. 
It belongs to the class of Bayesian methods. 
One variant has been discussed in detail in~\cite{Karpie:2019eiq}. We follow a slightly different
procedure here (see also~\cite{Young:2019}). Our model assumption is the general form of a valence quark type distribution
\begin{equation}
p^{\rm val}(x,a,b,c) = \frac{a \,x^{b}\, (1-x)^c \,\Gamma (b+c+3)}{\Gamma (b+2) \Gamma (c+1)}\,.
      \label{pval}
\end{equation}
We can determine the Compton amplitude from (\ref{opess2}) analytically (for $\omega < 1$)
\begin{eqnarray}
 T_{33}^{\rm val}(\omega)&=& 2^{-b-c-1}\, \sqrt{\pi }\, a \,\omega ^2 \, \Gamma (b+c+3) \times\nonumber\\ 
   & & _3\tilde{F}_2\left(1,\frac{b+2}{2},\frac{b+3}{2};\frac{1}{2} (b+c+3),\frac{1}{2}
   (b+c+4);\omega ^2\right)\\
   \label{Tanalytic}
   &=& c_1(a,b,c)\, \omega^2 + c_3(a,b,c)\, \omega^4 + c_5(a,b,c)\, \omega^6 + \dots\,,
   \label{Tanalyticser}
\end{eqnarray}
where $_3\tilde{F}_2$ is a regularized hypergeometric function. The power expansion of 
$T_{33}^{\rm val}(\omega)$ is given in (\ref{Tanalyticser}). We proceed by first generating
$N_{MC}$ Monte Carlo sets 
of model parameters $\{a,b,c\}_{k=1,\dots,N_{MC}}$. With these sets  the  quadratic deviations $\chi^2_k$
\begin{equation}
 \chi^2_k = \sum_{n,j}\,\left(T_{33,n}-T_{33,(k)}^{\rm val}(\omega_n)\right)\,C^{-1}_{nj}\,
 \left(T_{33,j}-T_{33,(k)}^{\rm val}(\omega_j)\right)
 \label{chi2}
\end{equation}
are computed. $T_{33,n}$ are the data for $\omega_n$, whereas $T_{33,(k)}^{\rm val}(\omega_n)$ 
is (2.5)
%%%(\ref{Tanalytic})
calculated for one triple $\{a,b,c\}_{k}$ at $\omega_n$.  $C^{-1}_{nj}$ is
the inverse covariance matrix of the data. The set $\chi^2_k$ is used
to make a weighted random choice out of the total set $\{a,b,c\}_k$ by the likelihood $exp(-\chi^2/2)$. 
This constitutes our sample 
parameter set from which we compute the means and the quantiles. Also in this case the model input
is crucial: the final values are inside the MC sets and the $\chi^2_k$ should contain reasonable small
minimal values. We call this method a Bayesian Monte Carlo (BMC) approach.
The resulting distribution is shown in the right panel of Fig. \ref{fig:T33dat}. The initial values 
of the parameters are selected uniformly distributed around some suitable values.
An analogous procedure can be used to determine the moments via relation (\ref{poly}). In this case 
the moments $t_j$ play the role of the parameters and are obtained directly from this approach.

Summarizing the SVD and the BMC approaches we favor the latter, because we recognize in the SVD solution 
oscillations around the exact result although we use ideal mock data. For real lattice data which are far more
scattered and which often have more significant uncertainties the SVD inversion gives very
unstable results.

\section{First results from lattice data}

\begin{figure}[htb]
%\hspace{0.06\textwidth}
  \begin{subfigure}[t]{0.47\textwidth}
    \includegraphics[width=\textwidth]{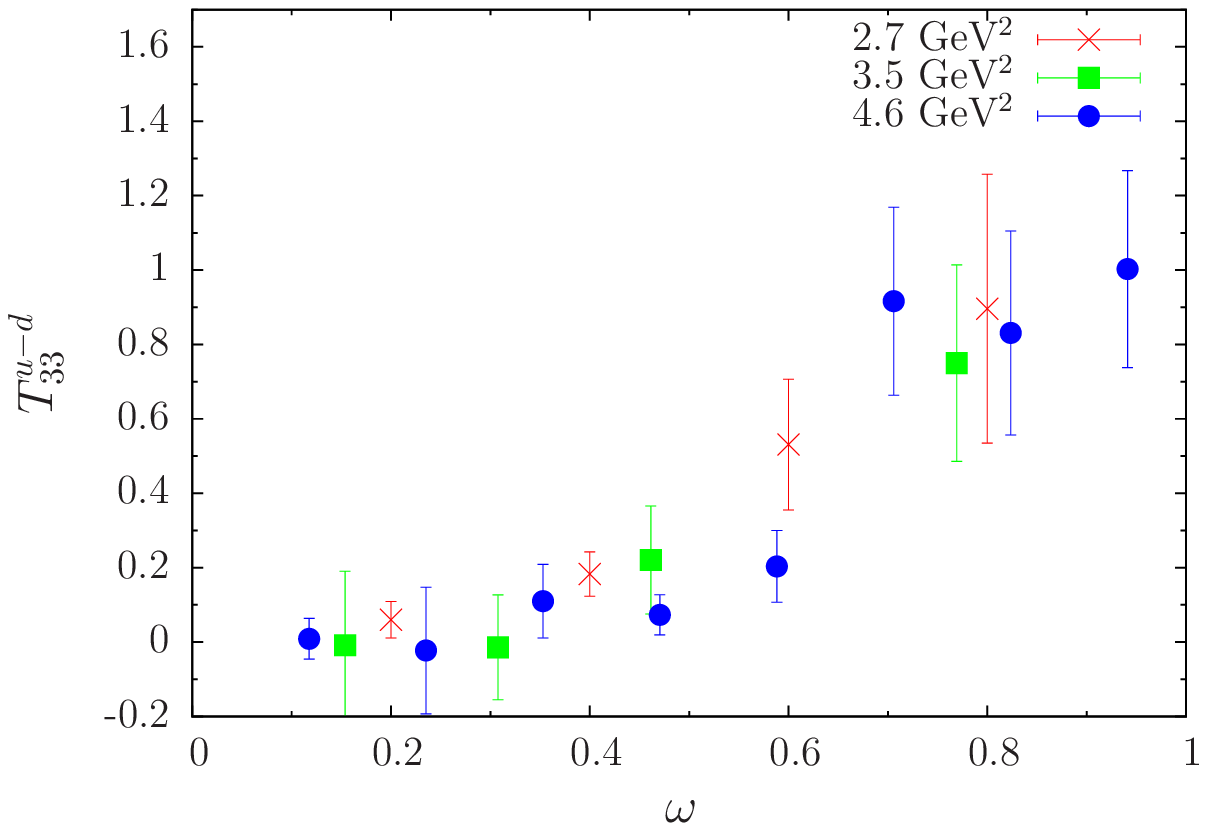}
  \end{subfigure}
  \hspace{0.04\textwidth}
  \begin{subfigure}[t]{0.47\textwidth}
    \includegraphics[width=\textwidth]{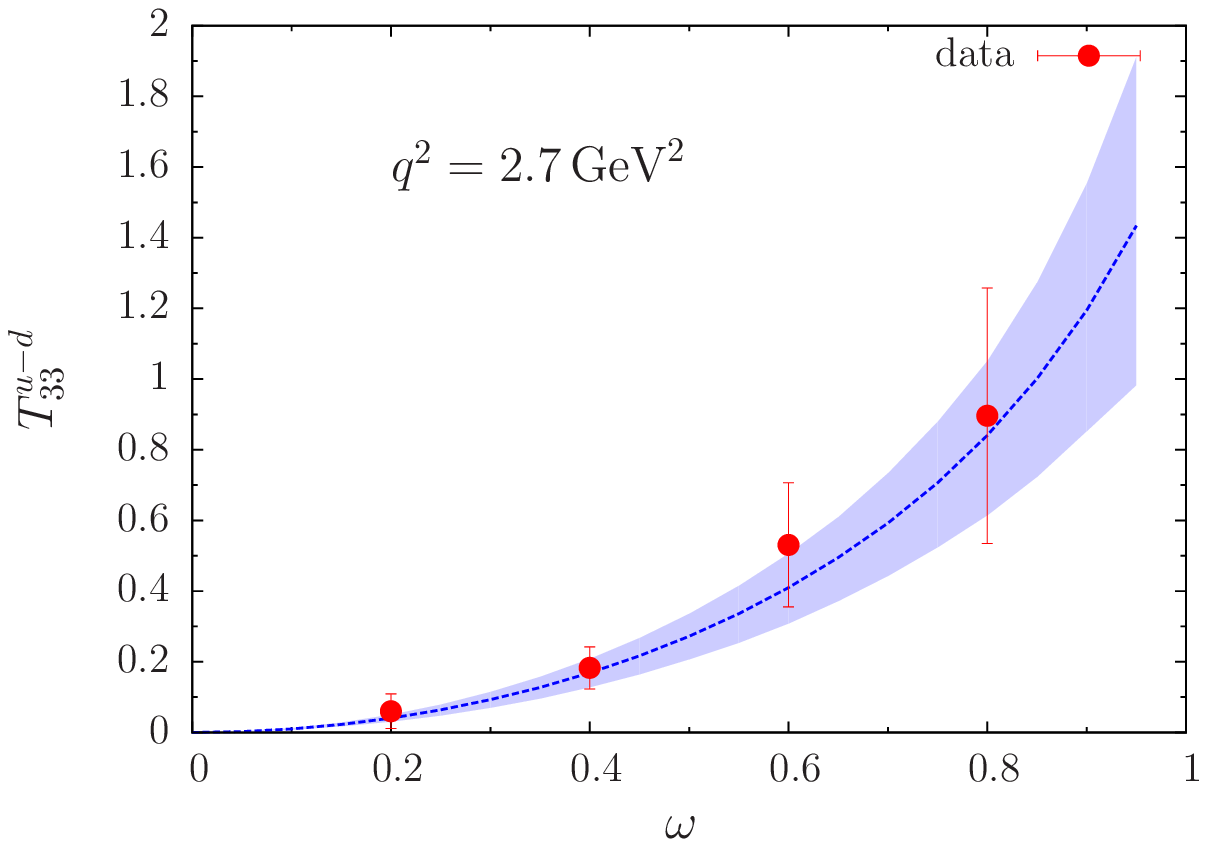}
  \end{subfigure}
%\vspace{-5mm}
\caption{Left: $T^{u-d}_{33}$ lattice data in the range $0 < \omega < 1$ for the three $q^2$ values
used for the analysis. Right: Result for (\protect\ref{T33mom}) together with the data for $q^2=2.7$ GeV$^2$.}
  \label{fig:t33latt}
\end{figure}

Now we investigate these methods with our latest lattice data for the nucleon Compton amplitude 
for the connected part of the combination $u-d$. We use $32^3\times 64 \,(\beta=5.5)$ lattices at the 
SU(3)--flavour symmetric
point $(\kappa_l=\kappa_s)$ and $M_\pi \approx 470$ GeV. 
In this paper only  data for $q^2 = 2.7,\, 3.5,\, 4.6\,$ GeV$^2$
are included. They are shown in the left panel of Fig. \ref{fig:t33latt}.

\begin{figure}[b]
%\hspace{0.06\textwidth}
  \begin{subfigure}[t]{0.47\textwidth}
    \includegraphics[width=\textwidth]{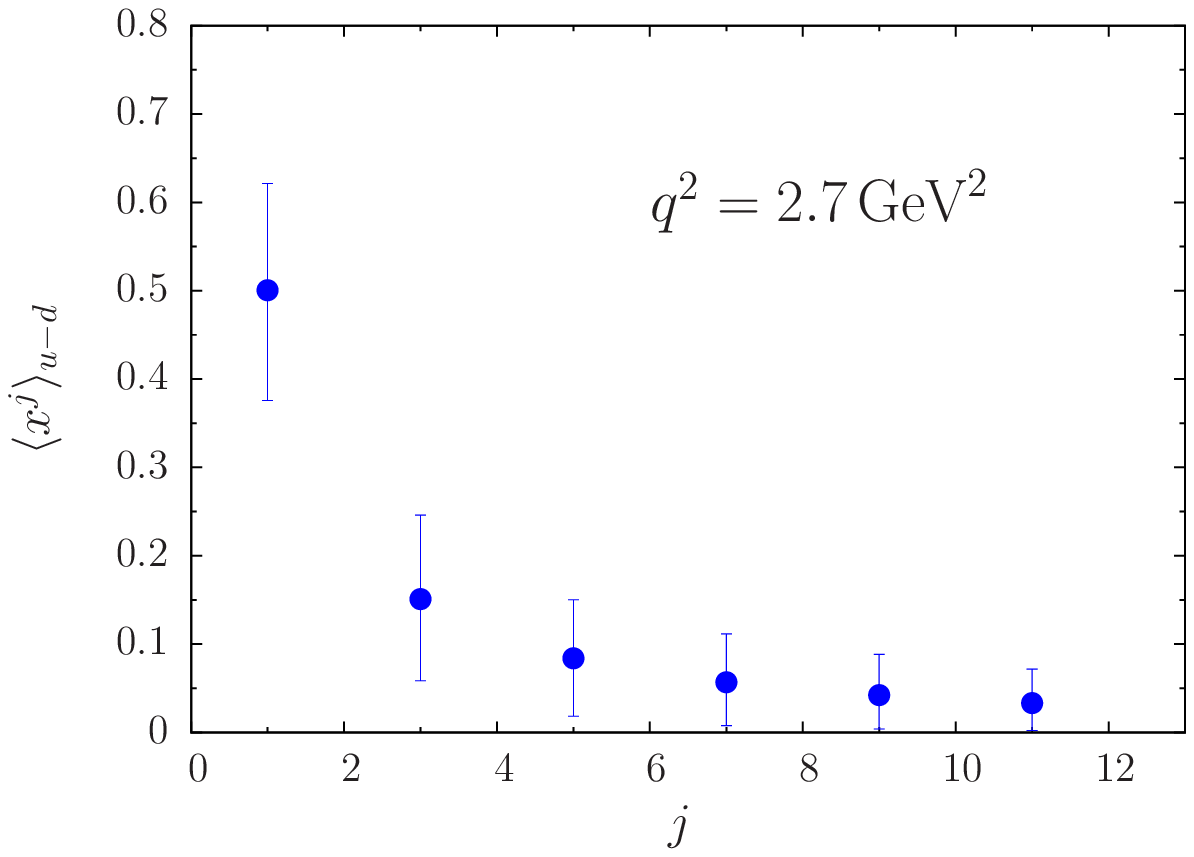}
  \end{subfigure}
  \hspace{0.04\textwidth}
  \begin{subfigure}[t]{0.47\textwidth}
    \includegraphics[width=\textwidth]{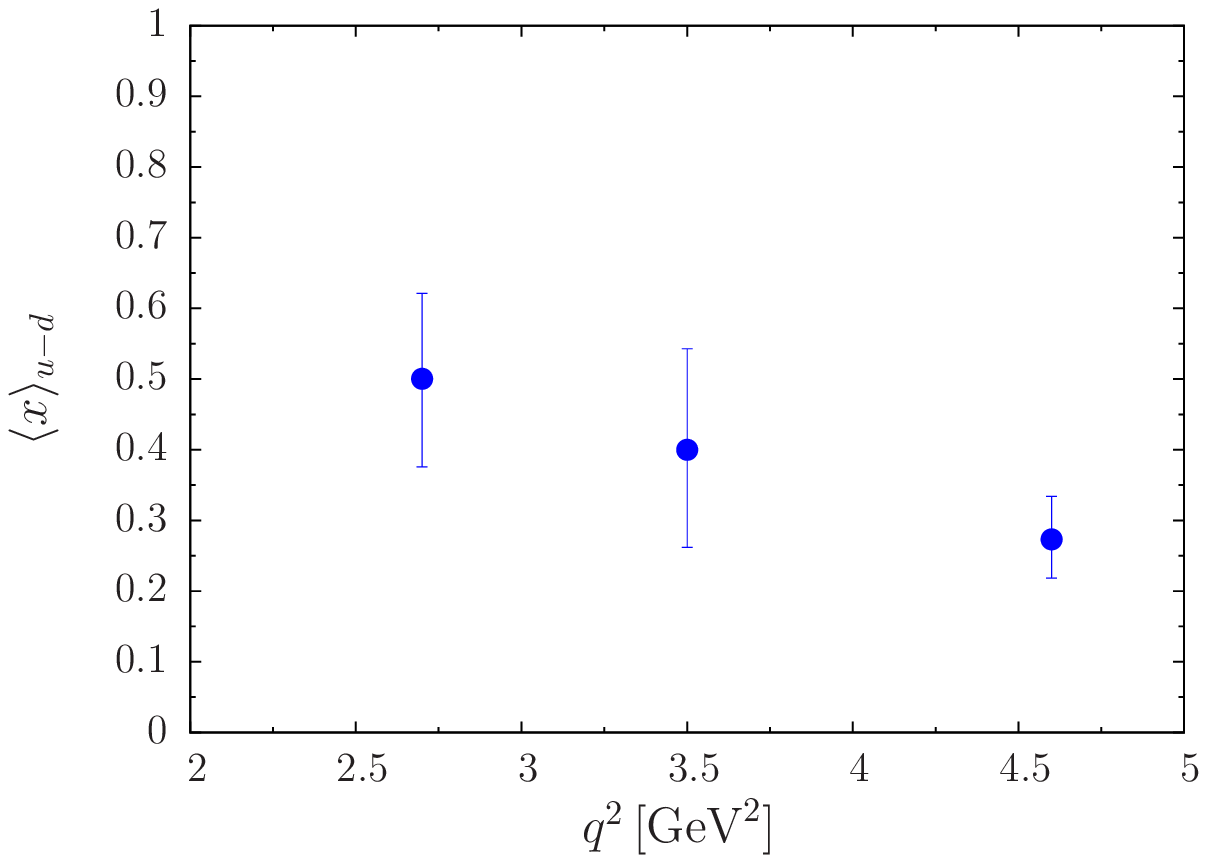}
  \end{subfigure}
%\vspace{-5mm}
\caption{Left: The first moments for $q^2=2.7$ GeV$^2$. The error bars are the quantiles encompassing
$68 \%$ of the data. 
Right: The first moment for $q^2=2.7,3.5$ and $4.6$ GeV$^2$.}
  \label{fig:momres}
\end{figure}

As a first step we determine the first moments using (\ref{poly}) as the defining relation.
In this paper we restrict ourselves to order $\omega^{12}$.
In order to get information about the $q^2$--dependence we apply our BMC procedure to each of the three 
data sets mentioned above. We compute the $\chi_k^2$ values from (\ref{chi2}), now with
\begin{equation}
 T_{33,(k)}^{\rm val}(\omega_n)=4\, \sum_{j=1}^6\,t^{(k)}_{2j-1}\,\omega_n^{2j} \,.
 \label{T33mom}
\end{equation}
We select $N_{MC}$ sets by sampling $\{a, b, c\}_k$ uniformly from intervals suggested by phenomenology
and
determine the moments $t(a,b,c)^{(k)}_i$ according to their valence quark beavior.
(A random selection with $t_1^{(k)} \ge t_3^{(k)} \ge \dots \ge 0$ as discussed in~\cite{Young:2019} leads to very similar results.)
We generate 100,000 MC data sets and from that we
select a subset of 500 samples weighted by the likelihood $exp(-\chi^2/2)$. From this subset we compute the $t_i$.
The resulting
Compton amplitude is given in the right panel of Fig. \ref{fig:t33latt} for $q^2 = 2.7 $ GeV$^2$ where we
observe a reasonable agreement with the data.
The moments themselves are presented in Fig. \ref{fig:momres}. They show their expected behavior 
with increasing order.
For the first
moment we observe a slight dependence on $q^2$.

In the same spirit we try to obtain the complete particle distribution function. As  data
we use the subset with $q^2=2.7$ GeV$^2$. 
Concerning the priors, we are guided by the success of the moments determination above. We
sample the first moment uniformly out of the interval $[0 \dots 1]$ and let the BMC method
compute it from to the lattice data. This is supported by 
our model ansatz  (\ref{pval}) since
\begin{equation}
\langle x \rangle = \int_0^1\, {\rm dx}\,\, x \,p^{\rm val}(x,a,b,c) = a \,.
      \label{pvalm}
\end{equation}
For the parameters $b$ and $c$ we choose input intervals suggested by phenomenology. Other prior
schemes will be investigated in a forthcoming paper.

\begin{figure}[t]
%\hspace{0.06\textwidth}
  \begin{subfigure}[t]{0.47\textwidth}
    \includegraphics[width=\textwidth]{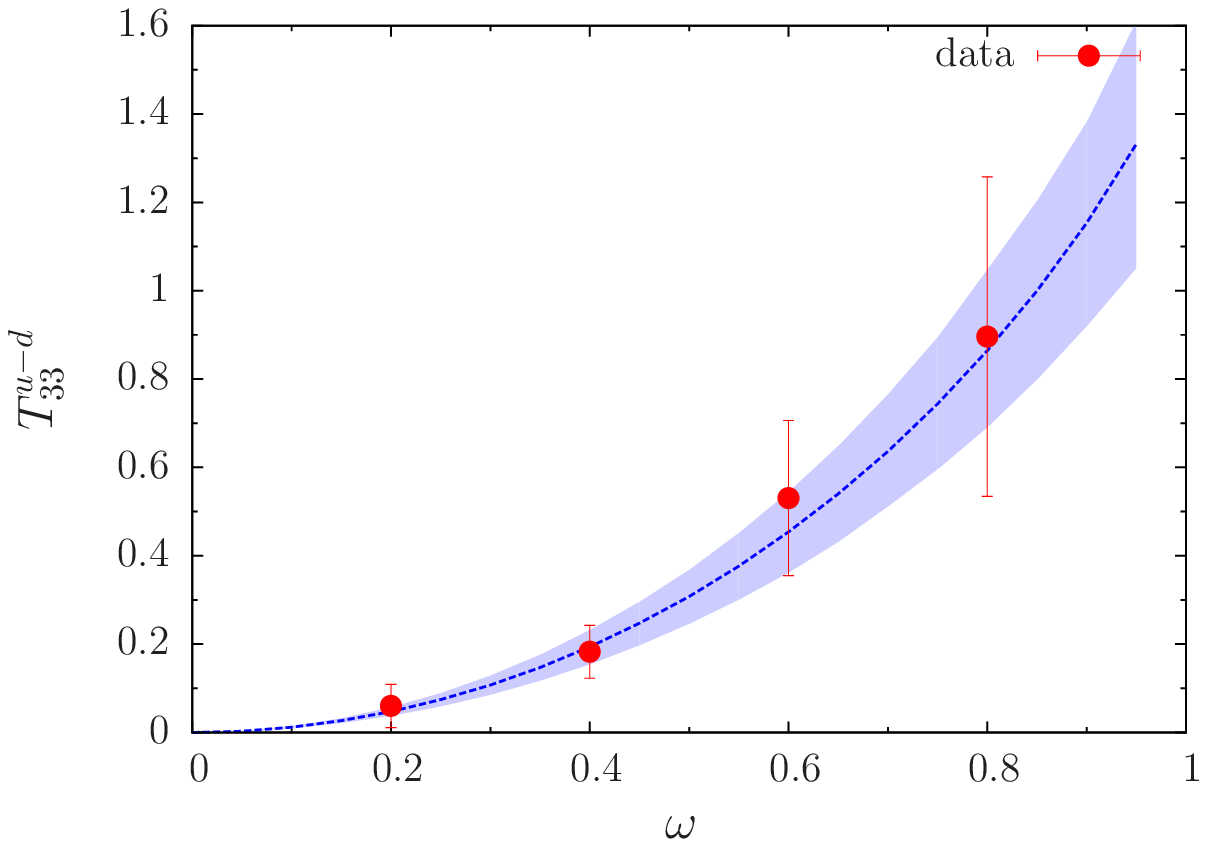}
  \end{subfigure}
  \hspace{0.04\textwidth}
  \begin{subfigure}[t]{0.47\textwidth}
    \includegraphics[width=\textwidth]{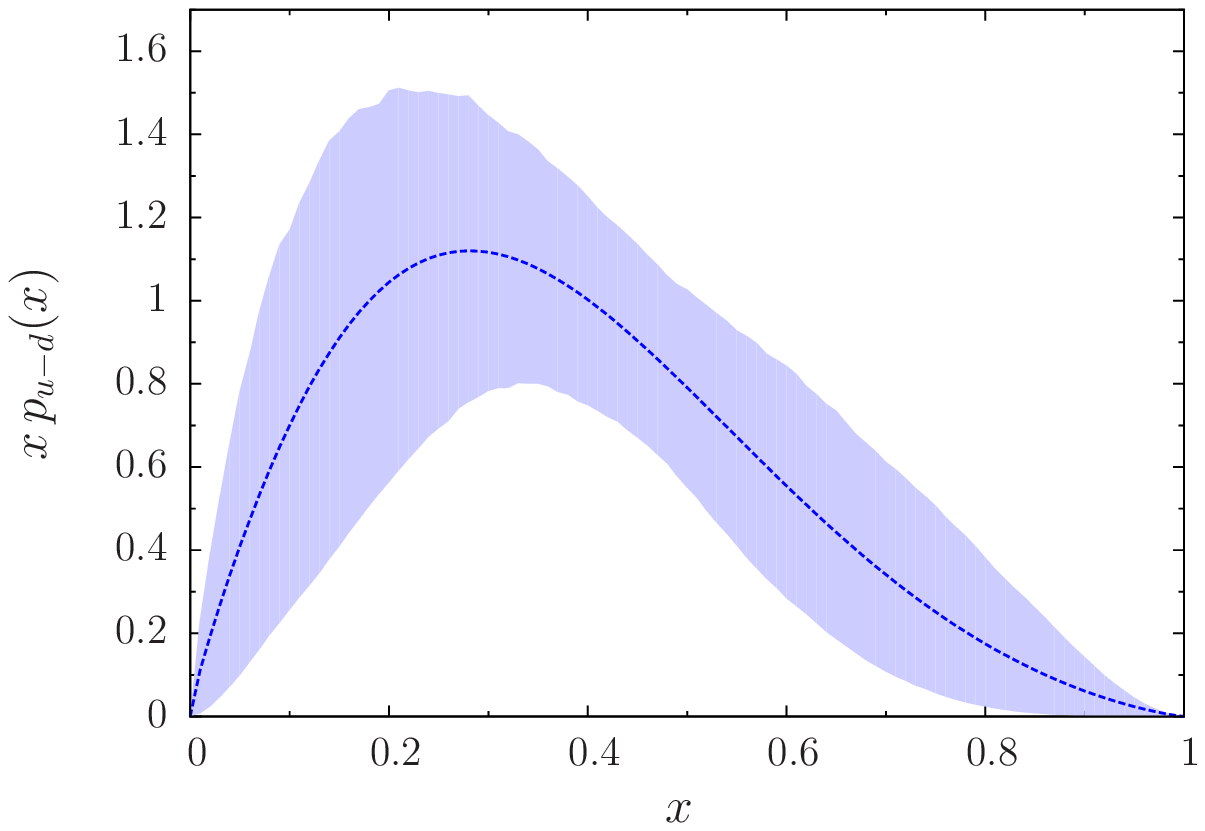}
  \end{subfigure}
%\vspace{-5mm}
\caption{Left: The Compton amplitude (\protect\ref{Tanalytic}) with parameters obtained with BMC together
with the data.
Right: The resulting valence type distribution function. The shaded area is the $68 \%$ quantile.}
  \label{fig:pvalres}
\end{figure}

\begin{figure}[!htb]
%\hspace{0.06\textwidth}
  \begin{subfigure}[b]{0.47\textwidth}
    \includegraphics[width=\textwidth]{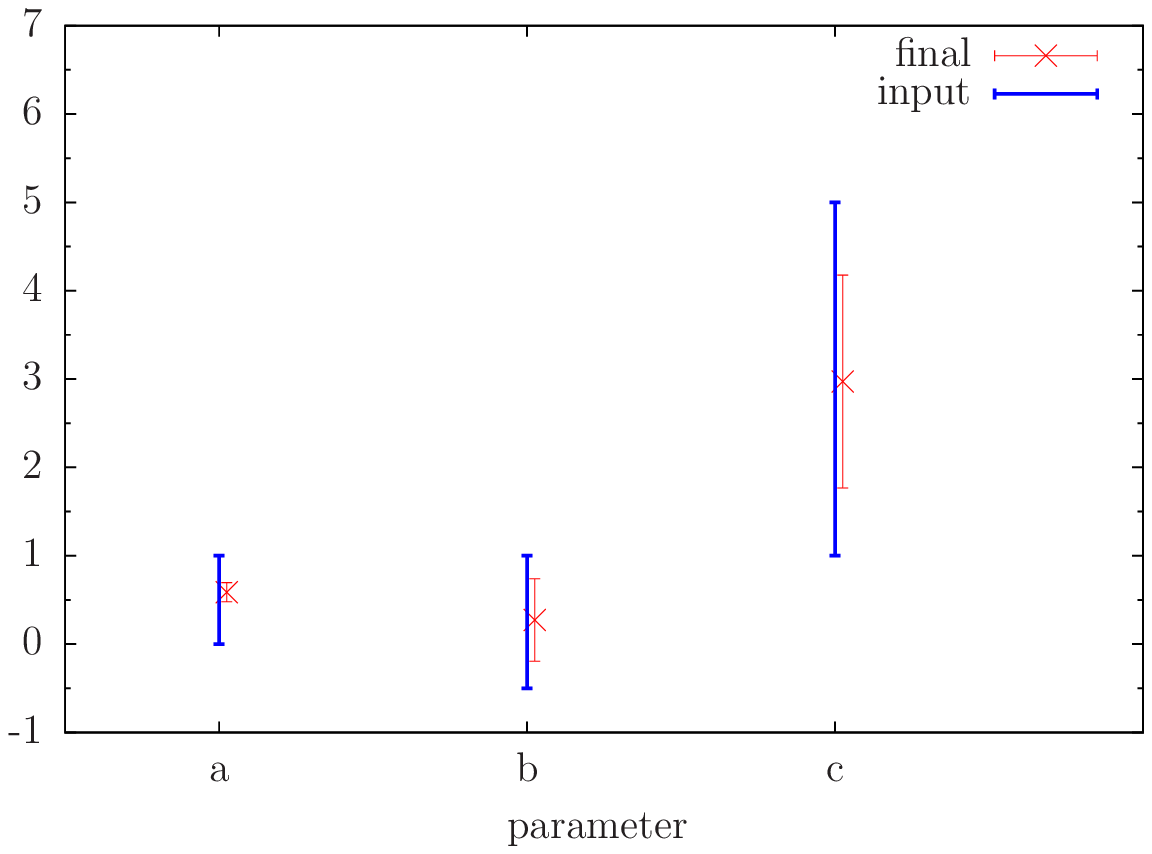}
  \end{subfigure}
  \hspace{0.04\textwidth}
  \begin{subfigure}[t]{0.47\textwidth}
    \includegraphics[width=\textwidth]{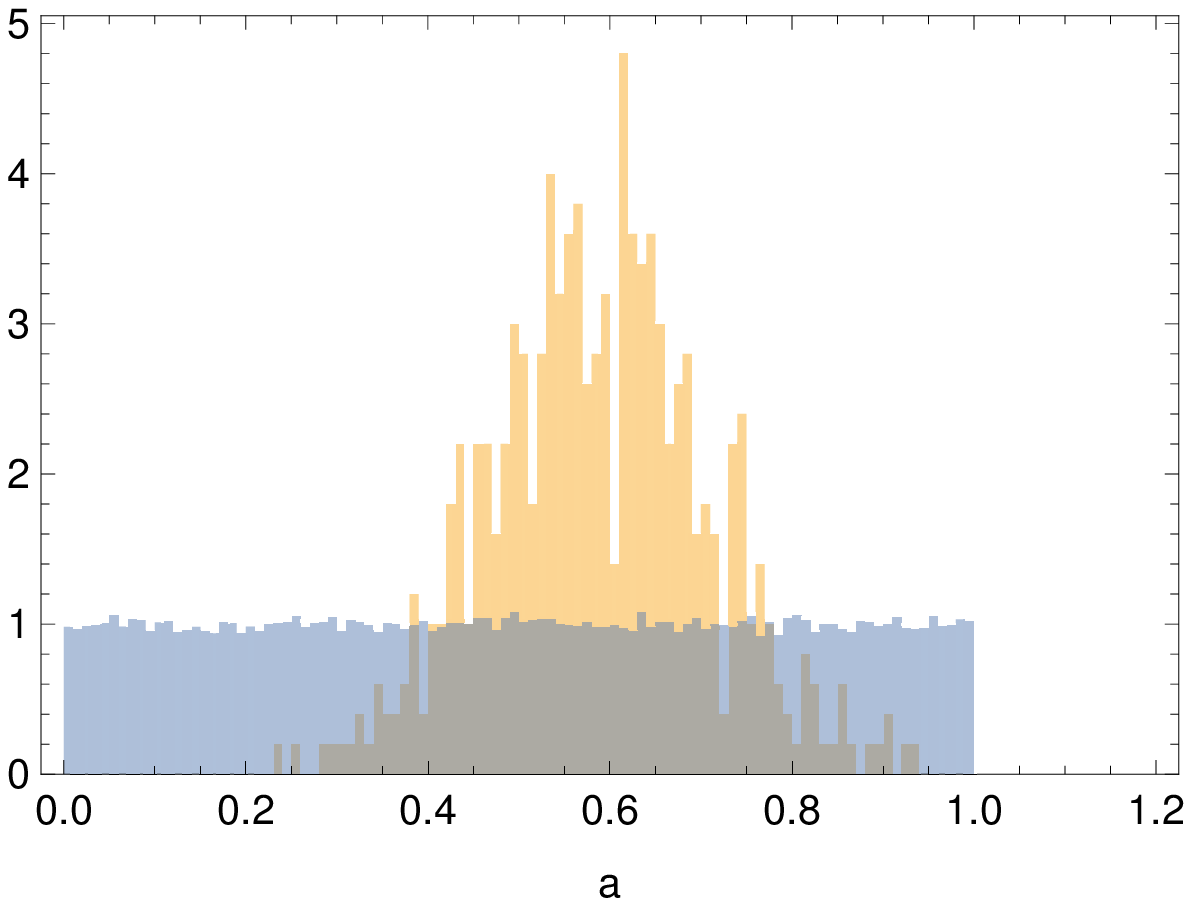}
  \end{subfigure}
%\vspace{-5mm}
\caption{Left: The parameters (a,b,c) of the valence distribution (\protect\ref{pval}). The blue lines are 
the input value ranges, the red data points result from the BMC approach. Again the error bars are the $68 \%$ quantiles.
Right: Histogram for parameter $a$. Blue: input parameter range. Yellow: parameter range of the sample according
to $\chi^2$ selection.}
  \label{fig:respar}
\end{figure}

We find using the mean curve and its quantile borders 
%%\begin{equation}
$ \int_0^1\, {\rm dx}\,\, x \,p^{\rm res}_{u-d}(x)~= 0.58^{+25}_{-26} \,$,
%%\end{equation}
consistent with the first moment given in
Fig. \ref{fig:momres}. Additionally, inserting the resulting mean values of the 
parameters in (\ref{Tanalyticser}) we
find $c_1 \approx 1.09$ -- also compatible with the moments.
The results are shown in Fig. \ref{fig:pvalres}. One recognizes a strong similarity of the left
panel in Fig. \ref{fig:pvalres} with the right panel of Fig. \ref{fig:t33latt} which proves
the consistency of both approaches.
In order to demonstrate the effect of the BMC procedure we show in Fig.~\ref{fig:respar} the change
of the parameters from the uniformly input values (blue) to the final values (red). The histogram in the 
right panel demonstrates the transition from uniform input to the peaked distribution triggered
by the $\chi_k^2$ values. One recognizes that the procedure does not influence very much the values 
of the parameters
$b$ and $c$ but significantly shrinks the range for parameter $a$ towards the first moment.

%{\bf Summary: }We have tested some methods to invert the Fredholm equation for the structure functions. Using mock data
%from phenomenological particle distribution functions they worked at least qualitatively. The results 
%from the SVD and the Bayesian methods show a reasonable behavior. Our current lattice data for $T_{33}$
%have still large errors. The model independent SVD approach is not suited, the solution is very unstable and
%oscillates in the best case of ideal data. Using
%the assumption of a valence form together with a smooth polynomial fit to $T_{33}$ 
%we get more or less stable results. We apply the BMC method for our lattice data and compute both
%the first moments and the distribution function. To get reliable results we use as input model the 
%form of a valence distribution suggested by the generation of the lattice data.

\section*{Acknowledgements}

%----------------------------------------------------------------------------

The numerical configuration generation (using the BQCD lattice
QCD program \cite{Haar:2017ubh})) and data analysis
(using the Chroma software library \cite{Edwards:2004sx}) was carried out
on the IBM BlueGene/Q and HP Tesseract using DIRAC 2 resources
(EPCC, Edinburgh, UK), the IBM BlueGene/Q (NIC, J\"ulich, Germany)
and the Cray XC40 at HLRN (The North-German Supercomputer
Alliance), the NCI National Facility in Canberra, Australia
(supported by the Australian Commonwealth Government)
and Phoenix (University of Adelaide).
RH was supported by STFC through grant ST/P000630/1.
HP was supported by DFG Grant No. PE 2792/2-1.
PELR was supported in part by the STFC under contract ST/G00062X/1.
GS was supported by DFG Grant No. SCHI 179/8-1.
RDY and JMZ were supported by the Australian Research Council Grants
DP140103067 and DP190100297.
We thank all funding agencies

\end{document}